\documentclass[a4paper,twoside]{article}

\usepackage{epsfig}
\usepackage{subfigure}
\usepackage{calc}
\usepackage{amssymb,amsthm,amstext}
\usepackage{bm}
\usepackage{amsmath,amsfonts,graphicx}
\usepackage{epstopdf}
\usepackage{multicol}
\usepackage{pslatex}
\usepackage{apalike}
\usepackage{listings}
\usepackage{mathptmx}
\usepackage{SCITEPRESS}
\usepackage{setspace}
\usepackage{algorithm}
\usepackage{algorithmic}
\usepackage{mathrsfs}
\usepackage{titlesec}
\titlespacing*{\section} {1pt}{3ex plus 1ex minus .2ex}{2.3ex plus .2ex}

\usepackage[small]{caption}
\newcommand{\tabincell}[2]{\begin{tabular}{@{}#1@{}}#2\end{tabular}}%
\makeatletter
\def\hlinewd#1{%
  \noalign{\ifnum0=`}\fi\hrule \@height #1 \futurelet
   \reserved@a\@xhline}
\makeatother
\begin{document}

\title{Multi-sensor authentication to improve smartphone security}

\author{\authorname{Wei-Han Lee, Ruby B. Lee}
\affiliation{Princeton Architecture Lab for Multimedia and Security (PALMS)\\Department of Electrical Engineering, Princeton University, Princeton, NJ, USA}
\email{\{weihanl, rblee\}@princeton.edu}
}


\abstract{The widespread use of smartphones gives rise to new security and privacy concerns. Smartphone thefts account for the largest percentage of thefts in recent crime statistics. Using a victim's smartphone, the attacker can launch impersonation attacks, which threaten the security of the victim and other users in the network. Our threat model includes the attacker taking over the phone after the user has logged on with his password or pin. Our goal is to design a mechanism for smartphones to better authenticate the current user, continuously and implicitly, and raise alerts when necessary. In this paper, we propose a multi-sensors-based system to achieve continuous and implicit authentication for smartphone users. The system continuously learns the owner's behavior patterns and environment characteristics, and then authenticates the current user without interrupting user-smartphone interactions. Our method can adaptively update a user's model considering the temporal change of user's patterns. Experimental results show that our method is efficient, requiring less than 10 seconds to train the model and 20 seconds to detect the abnormal user, while achieving high accuracy (more than 90\%). Also the combination of more sensors provide better accuracy. Furthermore, our method enables adjusting the security level by changing the sampling rate.}
\onecolumn \maketitle \normalsize \vfill
\section{\uppercase{Introduction}}
\label{Second:introduction}
\noindent In recent years, we have witnessed an increasing development of mobile devices such as smartphones and tablets. Smartphones are also becoming an important means for accessing various online services, such as online social networks, email and cloud computing. Many applications and websites allow users to store their information, passwords, etc. Users also save various contact information, photos, schedules and other personal information in their smartphones. No one wants personal and sensitive information to be leaked to others without their permission. However, the smartphone is easily stolen, and the attacker can have access to the personal information stored in the smartphone. Furthermore, the attacker can steal the victim's identity and launch impersonation attacks in networks, which would threaten the victim's personal and sensitive information like his bank account, as well as the security of the networks, especially online social networks. Therefore, providing reliable access control of the information stored on smartphones, or accessible through smartphones, is very important. But, first, it is essential to be able to authenticate the legitimate user of the smartphone, and distinguish him or her from other unauthorized users.

Passwords are currently the most common way for authentication. However, they suffer from several weaknesses. Passwords are vulnerable to attacks because they are easily guessed. They suffer from social engineering attacks, like phishing, pretexting, etc. The usability issue is also a serious factor, since users do not like to have to enter, and reenter, passwords or pins. A study \cite{consumer} shows that 64\% of users do not use passwords as an authentication mechanism on their smartphones. Hence, this paper proposes a means of implicit and continuous authentication, beyond the initial authentication by password, pin or biometric (e.g., fingerprint).

Implicit authentication does not rely on the direct involvement of the user, but is closely related to his/her biometric behavior, habits or living environment. We propose a form of implicit authentication realized by building the user's profile based on measurements from various sensors in a typical smartphone. Specifically, the sensors within the smartphones can reflect users' behavior patterns and environment characteristics. The recent development and integration of sensor technologies in smartphones, and advances in modeling user behavior create new opportunities for better smartphone security. 

\begin{table}[!t]
\caption{Sensors enabled in some popular smartphones.}
\begin{tabular}{|c|c|c|c|}
\hline
Sensor        & Nexus 5 & iphone 5s & Galaxy S5 \\ \hline
accelerometer & Yes & Yes & Yes \\ \hline
gyroscope     & Yes & Yes & Yes \\ \hline
magnetic field& Yes & Yes & Yes \\ \hline
light         & Yes & Yes & Yes \\ \hline
proximity     & Yes & Yes & Yes \\ \hline
pressure      & Yes & No  & Yes \\ \hline
temperature   & No  & No  & No \\ \hline
orientation   & Yes & No  & No \\ \hline
GPS           &Yes  &Yes  &Yes \\ \hline
MIC           &Yes  &Yes  &Yes \\ \hline
camera        &Yes  &Yes  &Yes \\ \hline
Network       &Yes  &Yes  &Yes \\ \hline
\end{tabular}
\label{sensors}
\end{table}

In this paper, we propose a multi-sensor-based system to achieve continuous and implicit authentication for smartphone users. The system leverages data collected by three sensors: accelerometer, orientation sensor, and magnetometer, in a smartphone, and then trains a user's profile using the SVM machine learning technique. The system continuously authenticates the current user without interrupting user-smartphone interactions. The smartphone's security system is alerted once abnormal usage is detected by our implicit authentication mechanism, so that access to sensitive information can be shut down or restricted appropriately, and further checking and remediation actions can be taken. Our authentication mechanism can adaptively update a user's profile every day considering that the user's pattern may change slightly with time. Our experimental results on two different data sets show the effectiveness of our proposed idea. It only takes less than 10 seconds to train the model everyday and 20 seconds to detect abnormal usage of the smartphone, while achieving high accuracy (90\%, up to 95\%). 

\begin{table*}[!t]
\caption{Sensor measurements, common usage and whether applications need the user's permission to access measurements.}
\begin{tabular}{|c|c|c|c|}
\hline
Sensor        & Description & Common Use & Permission \\ \hline
accelerometer & \tabincell{c}{Measures the acceleration force in $m/s^2$\\ on all three physical axes (x, y, and z)} & Motion detection & No \\ \hline
orientation   & \tabincell{c}{Measures degrees of rotation in rad\\ on all three physical axes (x, y, z)} & Rotation detection. & No \\ \hline
magnetometer& \tabincell{c}{Measures the ambient geomagnetic field \\for all three physical axes (x, y, z) in $\mu$T} & \tabincell{c}{Environment detection\\ (compass)} & No \\ \hline
gyroscope     & \tabincell{c}{Measures a device's rate of rotation in rad/s\\ on all three physical axes (x, y, and z)} & Rotation detection & No \\ \hline
light         & \tabincell{c}{Measures the ambient light level in lx} & \tabincell{c}{Environment detection} & No \\ \hline
proximity     & \tabincell{c}{Measures the proximity of an object in cm\\ relative to the view screen of a device} & Phone position during a call. & No \\ \hline
pressure      & \tabincell{c}{Measures the ambient air pressure in hPa} & Environment detection & No \\ \hline
temperature   & \tabincell{c}{Measures the ambient room temperature \\in degrees Celsius} & Environment detection & No \\ \hline
GPS           & Positioning &Tracking and Positioning  &Yes \\ \hline
microphone    & Record voice  & Speech recognition  &Yes \\ \hline
camera        & Record image  &Face recognition  &Yes \\ \hline
network       & Provide user connection to internet & \tabincell{c}{Connectivity, location,\\ surfing patterns} &Yes \\ \hline
\end{tabular}
\label{permission}
\end{table*}
We arrived at our three-sensor solution by first testing the performance on a single-sensor-based system, considering each of the accelerometer, the orientation sensor and the magnetometer. We found that the authentication accuracy for measurements from the orientation sensor alone is worse than that of the accelerometer alone or the magnetometer alone. Then, we test a two-sensors-based system, using pairwise combinations from these three sensors. This showed that the combination of multiple sensors can improve the accuracy of the resulting authentication. We then combined the measurements from all three sensors, and showed that while there was a slight performance improvement, this incremental improvement is much less than going from one to two sensors, and the authentication accuracy is already above 90\%, reaching 95\%. We also show that our method allows the users to adjust their security levels by changing the sampling rate of the collected data. 

The main contributions of our paper are summarized below.
\begin{itemize}
    \item We propose a multi-sensor-based system to achieve continuous and implicit authentication, which is accurate, efficient and flexible.
    \item We compare our three-sensor-based method with single-sensor and two-sensors-based methods on two real data sets. Our three-sensor-based method is shown to have the best performance.
    \item We also analyze the balance between the authentication accuracy and the training time. We give a reasonable
trade-off with respect to the sampling rate and the data size, that is practical and meaningful in the real world environment of commodity smartphone users.
\end{itemize}
\begin{spacing}{0.2}
\section{\uppercase{Background}}
\end{spacing}
\noindent 
\begin{table*}[] 
\caption{Comparison of our method (we just apply one of our results in PU data set.) and state-of-art research in implicit authentication (if the information is given in the paper cited, otherwise it is shown as n.a. (not available)). FP is false positive rate and FN is false negative rate. \emph{train} means the time for training the model and \emph{test} means the time for detecting the abnormal usage. The script column shows whether a user has to follow a script. If a script is required, we can not achieve implicit authentication without user participation.}\centering
\begin{tabular}{|c|c|c|c|c|c|c|}
\hline
            &Devices& Sensors   & Method    & Accuracy &\tabincell{c}{ Detecting \\time} & \tabincell{c}{Script} \\ \hline
Our method   & \tabincell{c}{Nexus 5\\Android}& \tabincell{c}{orientation, \\magnetometer,\\accelerometer} & SVM & \tabincell{c}{90.23\%} & \tabincell{c}{train:6.07s\\test:20s} & No \\ \hline
\tabincell{c}{Kayacik et al., 2014} & Android & \tabincell{c}{light,\\orientation, \\magnetometer,\\accelerometer}& \tabincell{c}{temporal \&\\spatial model} & n.a. & \tabincell{c}{train: n.a.\\test:$\geq$122s} & No\\ \hline
\tabincell{c}{ Zhu et al., 2013}& Nexus S & \tabincell{c}{orientation, \\magnetometer,\\accelerometer} & \tabincell{c}{n-gram\\language\\model} & 71.3\%  & n.a. & Yes\\ \hline
\tabincell{c}{ Buthpitiya et al.,2011} & n.a. & GPS & \tabincell{c}{n-gram model\\on location} & 86.6\% &\tabincell{c}{train:n.a.\\test:$\geq$30min}& No\\ \hline
\tabincell{c}{ Trojahn et al., 2013} & \tabincell{c}{HTC\\Desire} & screen & \tabincell{c}{keystroke \&\\handwriting} & \tabincell{c}{FP:11\% \\ FN:16\% } & n.a. & Yes \\ \hline
\tabincell{c}{ Li et al., 2013} & \tabincell{c}{Motorola\\Droid} & screen & sliding pattern& 95.7\% &\tabincell{c}{train: n.a.\\test:0.648s} & Yes \\ \hline
\tabincell{c}{ Nickel et al., 2012} &\tabincell{c}{Motorola\\Milestone} & accelerometer & K-NN &\tabincell{c}{FP:3.97\%\\FN:22.22\%} & \tabincell{c}{train:1.5min\\test:30s} & Yes \\ \hline
\end{tabular}
\label{table1}
\end{table*}

\subsection{Smartphone inputs and sensors}
A unique feature of a smartphone is that it is equipped with a lot of sensors. Table \ref{sensors} lists some common sensors
in some of the most popular smartphones. Table \ref{permission} lists the sensors' functionality, description of the measurements made, what it can be used for in terms of user or smartphone authentication, and whether Android permissions are required to read the sensor's measurements. 

Smartphone sensor information include measurements from an accelerometer, gyroscope, magnetometer, orientation sensor, ambient light, proximity sensor, barometric pressure and temperature. Other more privacy sensitive inputs include a user's location as measured by his GPS location, WLAN, cell tower ID and Bluetooth connections. Also privacy sensitive are audio and video inputs like the microphone and camera. The contacts, running apps, apps' network communication pattern, browsing history, screen on/off state, battery status and so on, can also help to characterize a user.
\subsection{Related work}
Table \ref{table1} summarizes and compares our work with past work on sensor-based authentication.

With the increasing development of mobile sensing technology, collecting many measurements through sensors in smartphones is now becoming not only possible, but quite easy through, for example, Android sensor APIs. Mobile sensing applications, such as the CMU MobiSens \cite{cc1}, run as a service in the background and can constantly collect sensors' information from smartphones. Sensors can be either hard sensors (e.g., accelerometers) that are physically-sensing devices or soft sensors that record information of a phone's running status (e.g., screen on/off). 

Continuous authentication on smartphones is likely to become an interesting new research area, given the easily accessible data today in smartphones. 

In \cite{cc3}, a lightweight, and temporally \& spatially aware user behavior model is proposed for authentication based on both hard and soft sensors. They considered four different attacks and showed that even the informed insider can be detected in 717 seconds. However, they did not quantitatively show the accuracy. In comparison, our method not only clearly shows high accuracy performance but also requires much less detection time (e.g., we only need 20 seconds to detect an abnormal user while training the profiles for less than 10 seconds.)

SenSec \cite{cc2} constantly collects data from the accelerometer, gyroscope and magnetometer, to construct the gesture model while the user is using the device. SenSec is shown to achieve an accuracy of 75\% in identifying users and 71.3\% in detecting the non-owners. However, they ask users to follow a script, i.e., a specific series of actions, for authentication. In comparison, we do not need users to follow a specific script while still getting good authentication accuracy, higher than 90\%.
\begin{figure*}[!t]
\centering
\includegraphics[width=6.2in,height=1.5in]{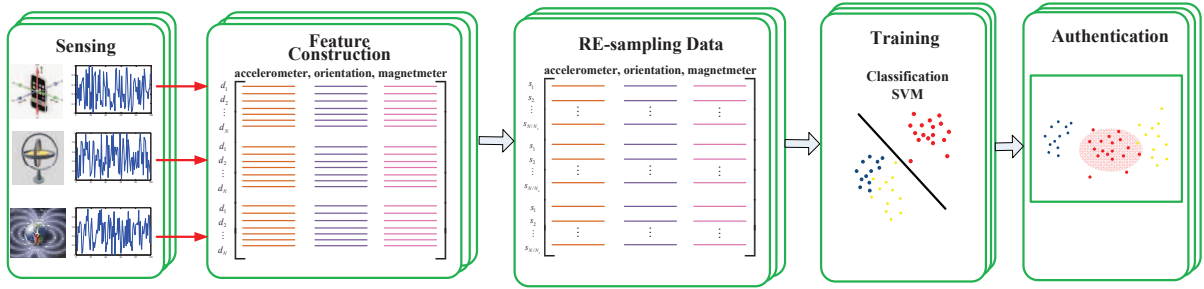}
\DeclareGraphicsExtensions.
\caption{In our method, we first construct a vector at each time by using sensors' data (In experiment, we use 9 values from the accelerometer, magnetometer and orientation sensor in a smartphone.)  After that, we re-sample the data collected from the sensors in a smartphone. Then, we train the re-sampled data with the SVM technique to get a user's profile. Based on the user's profile, we can do the implicit authentication.}
\label{mechanism}
\end{figure*}

In \cite{cc4}, an n-gram geo-based model is proposed for modeling a user's mobility pattern. They use the GPS sensor to demonstrate that the system could detect abnormal activities (e.g., a phone being stolen) by analyzing a user's location history, and the accuracy they achieve is 86.6\%. However, they just utilize a single sensor for authentication, which largely limits their performance. By exploiting multiple sensors, our method achieves better accuracy.

Biometric-based systems have also been used to achieve continuous and unobservable authentication for smartphones \cite{cc5} \cite{cc6} \cite{cc7}. However, they ask users to follow a script for authentication. In comparison, we do not need users to follow a specific script while still getting good authentication accuracy.
\cite{cc5} developed a mixture of a keystroke-based and a handwriting-based method to realize authentication through the screen sensor. Their approach has 11\% false acceptance rate and 16\% false rejection rate.
\cite{cc6} proposed another biometric method to do authentication for smartphones. They exploited five basic movements (sliding up, down, right, left and tapping) and the related combinations as the user's features, to perform authentication.
An accelerometer-based biometric gait recognition to authenticate smartphones through k-NN algorithm was proposed in \cite{cc7}. Their work is based on the assumption that different people have different walking patterns. Their process only takes 30 seconds. However, their approach asks the users to follow a script, where they just record the data when the user is walking. In comparison, we do not need the user to follow any script, which means that we can provide continuous protection without user interaction, while their approach can only guarantee security for walking users.

The fact that sensors reflect an individual's behavior and environment can not only be used for authentication, but can also lead to new attacks. \cite{cc8} proposed an attack to infer a user's input on a telephone key pad from measurements of the orientation sensor. They used the accelerometer to detect when the user is using a smartphone, and predicted the PIN through the use of gyroscope measurements.

Sensors also reflect environmental information, which can be used to reveal some sensitive information. By using measurements from an accelerometer on a smartphone to record the vibrations from a nearby keyboard \cite{cc11}, the authors could decode the context. In \cite{cc10}, the authors show that the gyroscope can record the vibration of acoustic signals, and such information can be used to derive the credit card number. 
\section{KEY IDEAS}
\noindent Some past work only consider one sensor to do authentication \cite{cc4}\cite{cc5}\cite{cc6}\cite{cc7}. We will show that the authentication accuracy can be improved by taking other sensors into consideration. We propose a multi-sensor-based technology with a machine learning method for implicit authentication, which only takes a short time to detect the abnormal user, but also needs less than 10 seconds to retrain the user's profile every day. First, we collect the data from the selected sensors. Then, we use the SVM technique as the classification algorithm to differentiate the usage patterns of various users and authenticate the user of the smartphone. 

Our methodology can be extended to other sensors in a straight-forward manner. Figure \ref{mechanism} shows our methodology, and the key ideas are presented below.

\subsection{Sensor Selection}
There are a lot of sensors built into smartphones nowadays as shown in Table \ref{sensors} and Table \ref{permission}. With smartphones becoming more connected with our daily lives, a lot of personal information can be stored in the sensors. The goal is to choose a small set of sensors that can accurately represent a user's characteristics. 
In this paper, we experiment with three sensors that are commonly found in smartphones: accelerometers, orientation sensors and magnetometers. They also represent different information about the user's behavior and environment: the accelerometer can detect coarse-grained motion of a user like how he walks \cite{cc7}, the orientation sensor can detect fine-grained motion of a user like how he holds a smartphone \cite{cc8}, and the magnetometer measurements can perhaps be useful in representing his environment. Furthermore, these sensors do not need the user's permission to be used in Android applications, which is useful for continuous monitoring for implicit authentication.

Also, these three sensors do not need the user to perform a sequence of actions dictated by a script--– hence facilitating implicit authentication. Note that our method is not limited to these three sensors, but can be easily generalized to different selections of hard or soft sensors, or to incorporate more sensors.
\subsection{Data Sets and Re-sampling}
\begin{spacing}{1}
We use two data sets, a new one which we collected locally by ourselves which we call the PU data set, and another data set which we obtained from the authors of a published paper \cite{cc3}, which we call the GCU data set. 

The PU data set is collected from 4 graduate students in Princeton University in 2014 based on the smartphone, Google Nexus 5 with Android 4.4. It contains sensor data from the accelerometer, orientation sensor and magnetometer with a sampling rate of 5 Hz. The duration of the data collected is approximately 5 days for each user. Each sensor measurement consists of three values, so we construct a vector from these nine values. We use different sampling rates as a factor in our experiments, to construct data points.

\setlength{\parskip}{0.2pt}
We use the second data set, called the GCU dataset version 2 \cite{cc3}, for comparison. This is collected from 4 users consisting of staff and students of Glasgow Caledonian University. The data was collected in 2014 from Android devices and contains sensor data from wifi networks, cell towers, application use, light and sound levels, acceleration, rotation, magnetic field and device system statistics. The duration of the data collected is approximately 3 weeks. For better comparison with our PU data set, we only use the data collected from the accelerometer, orientation sensor and magnetometer data. 

The sensor measurements originally obtained are too large to process directly. Hence, we use a re-sampling process to not only reduce the computational complexity but also reduce the effect of noise by averaging the data points. For example, if we want to reduce the data set by 5 times, we average 5 contiguous data points into one data point. In section 4, we will show that the time for training a user's profile can be significantly reduced by re-sampling. 
\end{spacing}
\subsection{Support Vector Machines}
\begin{figure}[!t]
\centering
\includegraphics[width=3in,height=1.8in]{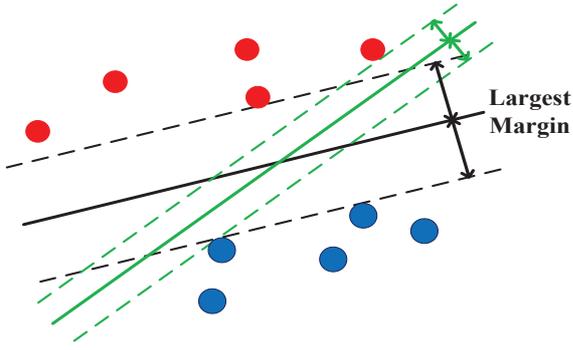}
\DeclareGraphicsExtensions.
\caption{Illustrating SVM. The purpose of SVM to find out the largest margin separating two groups of data }
\label{SVM_teach}
\end{figure}

The classification method used by prior work did not give very accurate results. Hence, we propose the use of the SVM technique for better authentication accuracy. 

Support Vector Machines (SVMs) are state-of-the-art large margin classifiers, which represent a class of supervised machine learning algorithms first introduced by \cite{svm2}. SVMs have recently gained popularity for human activity recognition on smartphones \cite{svm_recognization}. In this section, we provide a brief review of the related theory of SVMs \cite{svm}, \cite{svm2}.

After obtaining the features from sensors, we use SVM as the classification algorithm in the system. The training data is represented as $\mathcal{D}=\{(\bm{x}_i,\bm{y}_i)\in \mathcal{X}\times\mathcal{Y}:i=1,2,\dots ,n\}$ for $n$ data-label pairs. For binary classification, the data space is $\mathcal{X}=\mathbb{R}^d$ and the label set is $\mathcal{Y}=\{-1,+1\}$. The predictor $\bm{w}$ is $\mathcal{X}\rightarrow \mathcal{Y}$. The objective function is $J(\bm{w},\mathcal{D})$. The SVM finds a hyperplane in the training inputs to separate two different data sets such that the margin is maximized. Figure \ref{SVM_teach} illustrates the concept of SVM classification. A margin is the distance from the hyperplane to a boundary data point. The boundary point is called a support vector and there may exist many support vectors. The most popular method of training such a linear classifier is by solving a regularized convex optimization problem:
\begin{equation}
\bm{w}^{*}=\mathrm{argmin}_{\bm{w}\in\mathbb{R}^d}\frac{\lambda}{2}\|\bm{w}\|^2+\frac{1}{n}\sum_{i=1}^n l\left(\bm{w}, \bm{x}_i, y_i\right)
\label{eq1}
\end{equation}
where
\begin{equation}
l\left(\bm{w}, x, y\right)=\max\left(1-y\bm{w}^T \bm{x},0\right)
\label{eq2}
\end{equation}

\begin{figure*}[!t]
\centering
\subfigure[PU data set]{
\label{time_PU} 
\includegraphics[width=3.05in,height=2.1in]{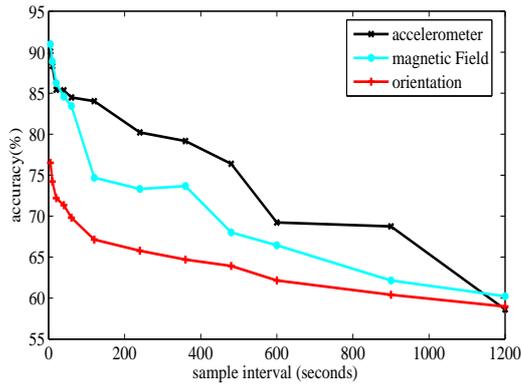}}
\subfigure[GCU data set]{
\label{time_GCU} 
\includegraphics[width=3.05in,height=2.1in]{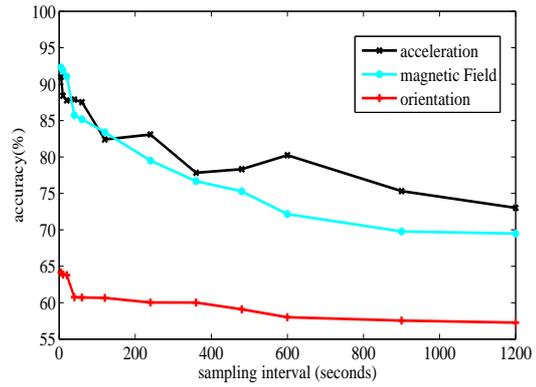}}
\caption{Authentication accuracy for single sensor system in (a) the PU data set, and (b) the GCU data set. Higher sampling rates give better accuracy for each sensor. The accelerometer and magnetometer have better performance than the orientation sensor. The reason is that both of them record a user's longer term characteristics, where the accelerometer somehow represents a user's walking style and the magnetometer records a user's general environment. However, the orientation sensor represents how the user holds a smartphone, which is more variable.}
\label{single} 
\end{figure*}
\begin{figure*}[!t]
\centering
\subfigure[PU data set]{
\label{time_PU} 
\includegraphics[width=3.05in,height=2.1in]{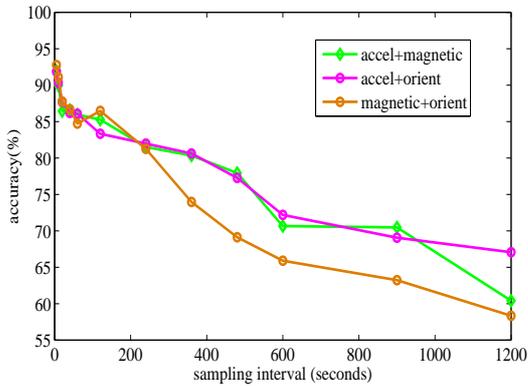}}
\subfigure[GCU data set]{
\label{time_GCU} 
\includegraphics[width=3.05in,height=2.1in]{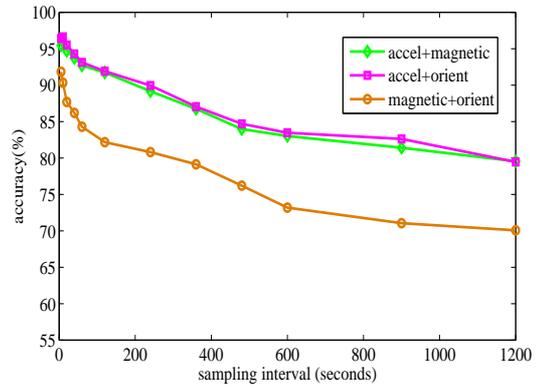}}
\caption{Authentication accuracy with SVM for a combination of two sensors, for (a) the PU data set, and (b) the GCU data set. The higher sampling rate gives better accuracy for each sensor.}
\label{double} 
\end{figure*}
\begin{figure*}[!t]
\centering
\subfigure[PU data set]{
\label{PU_all} 
\includegraphics[width=3.05in,height=2.1in]{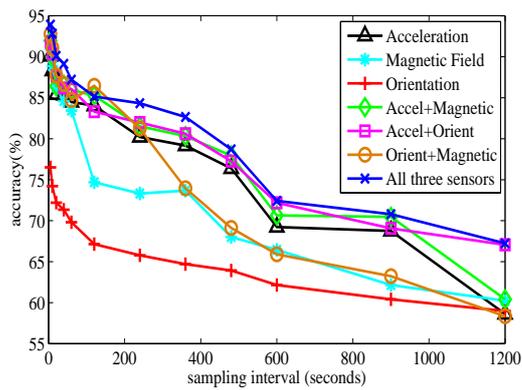}}
\subfigure[GCU data set]{
\label{GCU_all} 
\includegraphics[width=3.05in,height=2.1in]{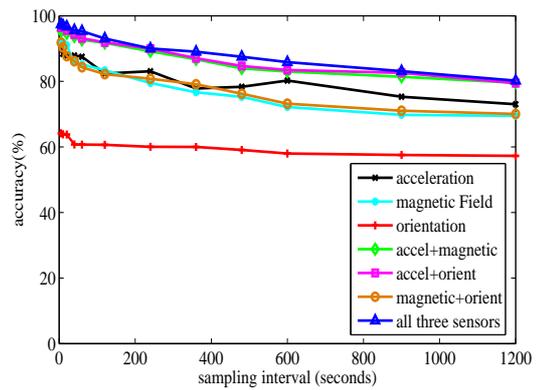}}
\caption{Authentication accuracy for single sensors, two sensors and three sensors, for the PU data set and the GCU data set. The higher sampling rate has better accuracy for each combination of sensors. Two sensors give better accuracy than using a single sensor, and three sensors further improves the accuracy.}
\label{all} 
\end{figure*}
\begin{table*}[!t]
\caption{The accuracy (\%) vs. sampling rate in both PU data set and GCU data set for all combinations of 1, 2 or 3 sensors.}
\begin{tabular}{|c|c|c|c|c|c|c|c|c|c|c|c|c|}
\hline
\tabincell{c}{sampling rate (s)} & 5& 10 & 20 & 40 & 60 & 120 & 240 & 360 & 480 &600 & 900 & 1200 \\ \hline
acc(PU)  & 90.1 & 88.3 & 85.4 & 85.3 & 84.5 & 84.0 & 80.2 & 79.2 & 76.4 & 69.2 & 68.8 & 58.6 \\ \hline
mag(PU) & 91.0 & 88.9 & 86.2 & 84.6 & 83.4 & 74.7 & 73.3 & 73.7 & 68.0 & 66.4 & 62.2 & 60.2 \\ \hline
ori(PU) & 76.5 & 74.2 & 72.2 & 71.3 & 69.8 & 67.1 & 65.8 & 64.7 & 63.9 & 62.1 & 60.4 & 59.0 \\ \hline
acc+mag(PU) & 92.0 & 90.0 & 86.4 & 86.6 & 85.9 & 85.3 & 81.5 & 80.3 & 77.9 & 70.6 & 70.5 & 60.4 \\ \hline
acc+ori(PU) & 91.8 & 90.3 & 87.7 & 86.2 & 86.1 & 83.3 & 82.0 & 80.6 & 77.3 & 72.2 & 69.1 & 67.1 \\ \hline
mag+ori(PU) & 92.8 & 91.1 & 87.7 & 86.7 & 84.7 & 86.5 & 81.3 & 74.0 & 69.1 & 65.9 & 63.2 & 58.3 \\ \hline
all(PU) & 93.9 & 92.8 & 90.1 & 89.1 & 87.2 & 85.2 & 84.3 & 82.7 & 78.7 & 72.4 & 70.8 & 67.2 \\ 
\hlinewd{0.3pt} \hline \hline \hlinewd{0.3pt} \hline
acc(GCU)  & 91.0 & 88.4 & 87.8 & 87.9 & 87.5 & 82.4 & 83.1 & 77.8 & 78.3 & 80.2 & 75.3 & 73.0\\ \hline
mag(GCU) & 92.3 & 91.2 & 91.0 & 85.7 & 85.2 & 83.4 & 79.5 & 76.7 & 75.3 & 72.2 & 69.8 & 69.5 \\ \hline
ori(GCU) & 64.2 & 63.9 & 63.8 & 60.8 & 60.7 & 60.6 & 60.0 & 60.0 & 59.1 & 58.0 & 57.5 & 57.3 \\ \hline
acc+mag(GCU) & 95.5 & 95.8 & 94.7 & 93.7 & 92.7 & 91.8 & 89.2 & 86.7 & 84.0 & 83.1 & 81.4 & 79.6 \\ \hline
acc+ori(GCU) & 96.4 & 96.6 & 95.5 & 94.3 & 93.1 & 92.0 & 90.0 & 87.1 & 84.7 & 83.5 & 82.7 & 79.4 \\ \hline
mag+ori(GCU) & 91.8 & 90.3 & 87.7 & 86.2 & 84.3 & 82.2 & 80.8 & 79.1 & 76.2 & 73.2 & 71.1 & 70.1 \\ \hline
all(GCU) & 97.4 & 97.1 & 96.7 & 95.7 & 95.3 & 93.1 & 90.0 & 89.1 & 87.5 & 85.9 & 83.1 & 80.2 \\ \hline
\end{tabular}
\label{num}
\end{table*}

The margin is $\frac{2}{||w||}$ in SVM. So, Equation \ref{eq1} minimizes the reciprocal of the margin (first part) and the misclassification loss (second part). The loss function in SVM is the Hinge loss (Equation \ref{eq2}) \cite{gentile1998linear}. Sometimes, we need to map the original data points to a higher dimensional space by using a kernel function so as to make training inputs easier to separate. In our classification, we label the smartphone owner's data as positive and all the other users' data as negative. Then, we exploit such a model to do authentication. Ideally, only the user who is the owner of the smartphone is authenticated, and any other user is not authenticated. In our experiments, we selected LIBSVM \cite{cc9} to implement the SVM. The input of our experiment is n positive points from the legitimate user and n negative data points from randomly selected n other users. The output is the user's profile for the legitimate user.
\begin{spacing}{0.5}
\section{EXPERIMENTAL RESULTS}
\end{spacing}
\noindent Figure \ref{mechanism} shows the steps in our experiments. The following are some settings in our experiments:
\begin{itemize}
    \item We use both the PU data set and the GCU data set.
    \item We use accelerometer, magnetometer and orientation sensors (can be extended to other sensors).
    \item We re-sample the data by averaging the original data, with the sampling rate changing from 1 second to 20 minutes.
    \item Each data is a 9-dimensional vector (three values for each sensor). We use SVM to train the data within one day to obtain a user's profile.
    \item We label one user's data as positive and the other users' data as negative, and randomly pick equivalent data from both positive and negative sets.
    \item We experiment with data from one sensor, a pair of two sensors, and all three sensors to train the user's profile. We show that multi-sensor-based authentication indeed improves the authentication accuracy.
    \item In our experiments, we use 10-fold cross validation, which means that the size of training data over the size of training data and testing data is 1/10.
\end{itemize}
\subsection{Single-sensor authentication}
From Figure \ref{single}, we observe the single-sensor-based system in both the PU data set and the GCU data set. First, we find that the accuracy increases with faster sampling rate because we use more detailed information from each sensor. Second, an interesting finding is that the accelerometer and the magnetometer have much better accuracy performance than the orientation sensor, especially for the GCU data set. We think this is because they both represent a user's longer-term patterns of movement (as measured by the accelerometer) and his general environment (as measured by the magnetometer). The orientation sensor represents how the user holds a smartphone \cite{cc8}, which may be more variable. Therefore, the accelerometer and magnetometer have better authentication accuracy. The difference is more marked in the GCU data set, but the overall relative accuracy of the three sensors is the same in both data sets. The accuracy is below 90\% even for fast sampling rates like 10 seconds (see also Table \ref{all}).
\subsection{Two-sensor authentication}
Fig. \ref{double} shows that for all pairwise combinations, accuracy increases with faster sampling rate. The combination of data from two sensors indeed gives better authentication accuracy than using a single sensor (see Table \ref{all}). The average improvement from one sensor to two sensors is 7.4\% in PU data set (14.6\% in GCU data set) when the sampling rate is 20 seconds. Another interesting finding is that using a combination of magnetometer and orientation sensors is worse than the other two pairs which include an accelerometer. In fact, the combination of magnetometer and orientation sensors is not necessarily better than using just the accelerometer (see also Table \ref{num}). Therefore, choosing good sensors is very important. Also, using higher sampling rate gives better accuracy.

\begin{figure*}[]
\centering
\subfigure[]{
\label{time_PU} 
\includegraphics[width=3.05in,height=2.15in]{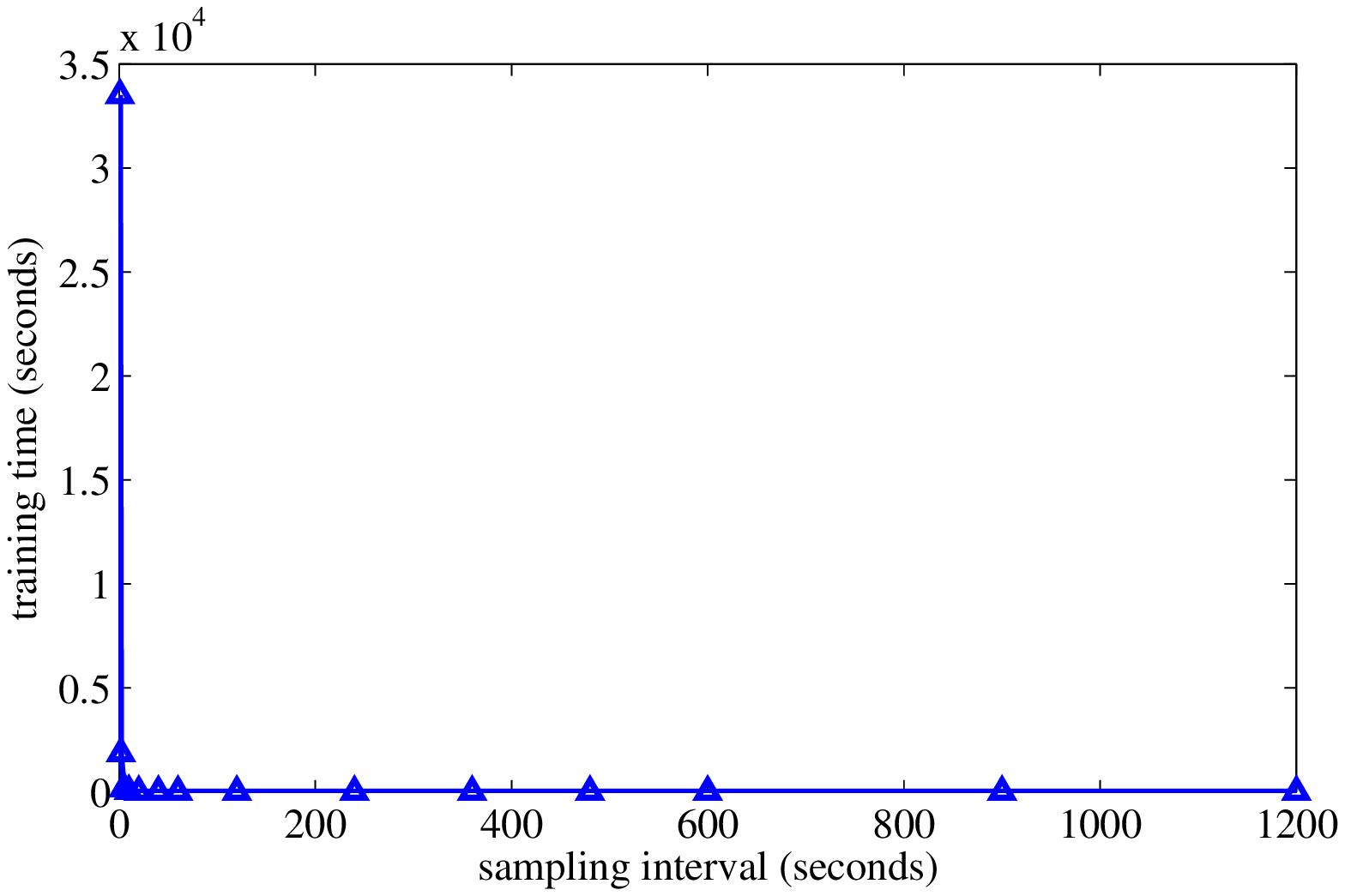}}
\subfigure[]{
\label{time_GCU} 
\includegraphics[width=3.05in,height=2.15in]{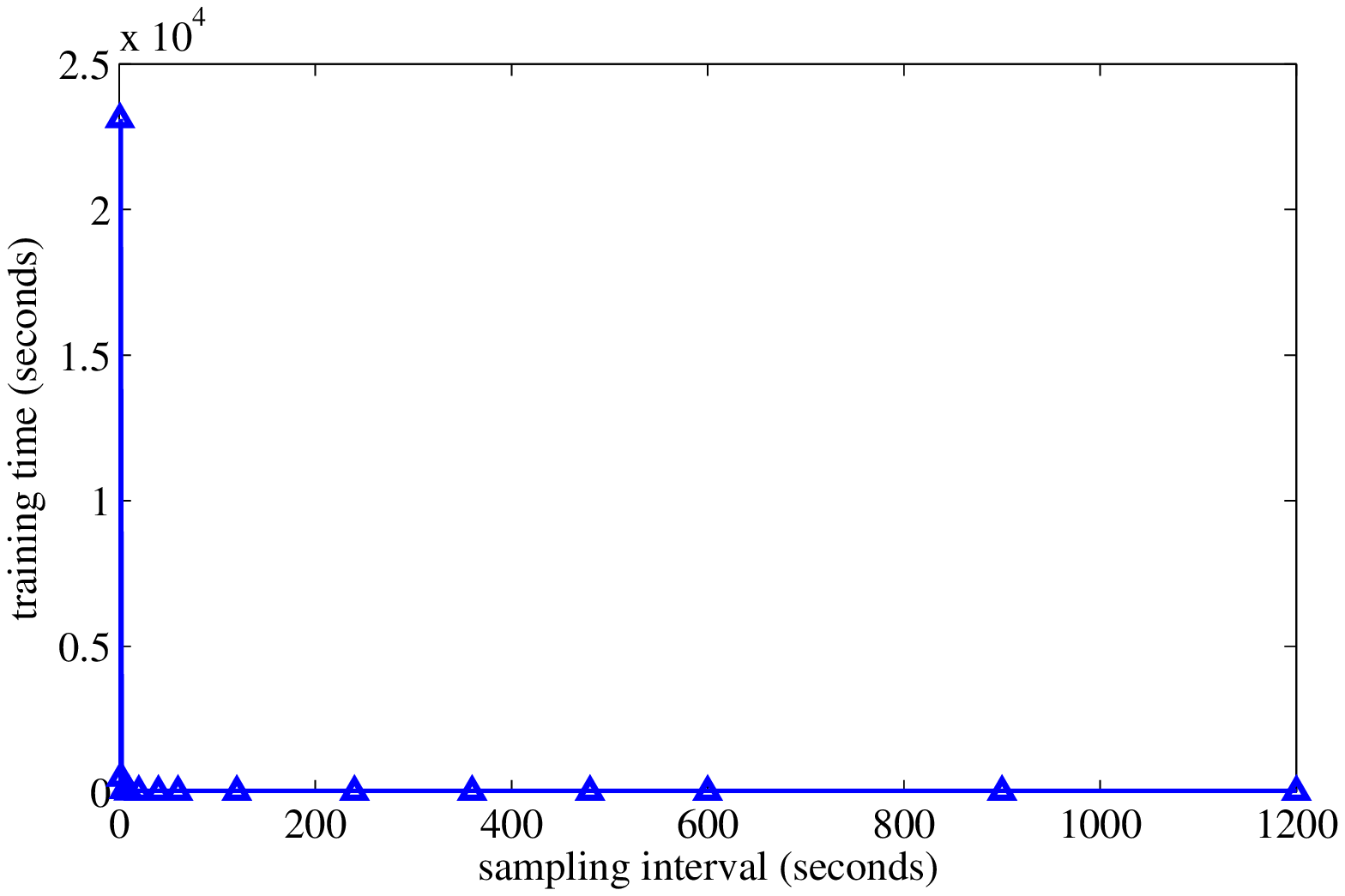}}
\subfigure[]{
\label{time_GCU} 
\includegraphics[width=3.05in,height=2.15in]{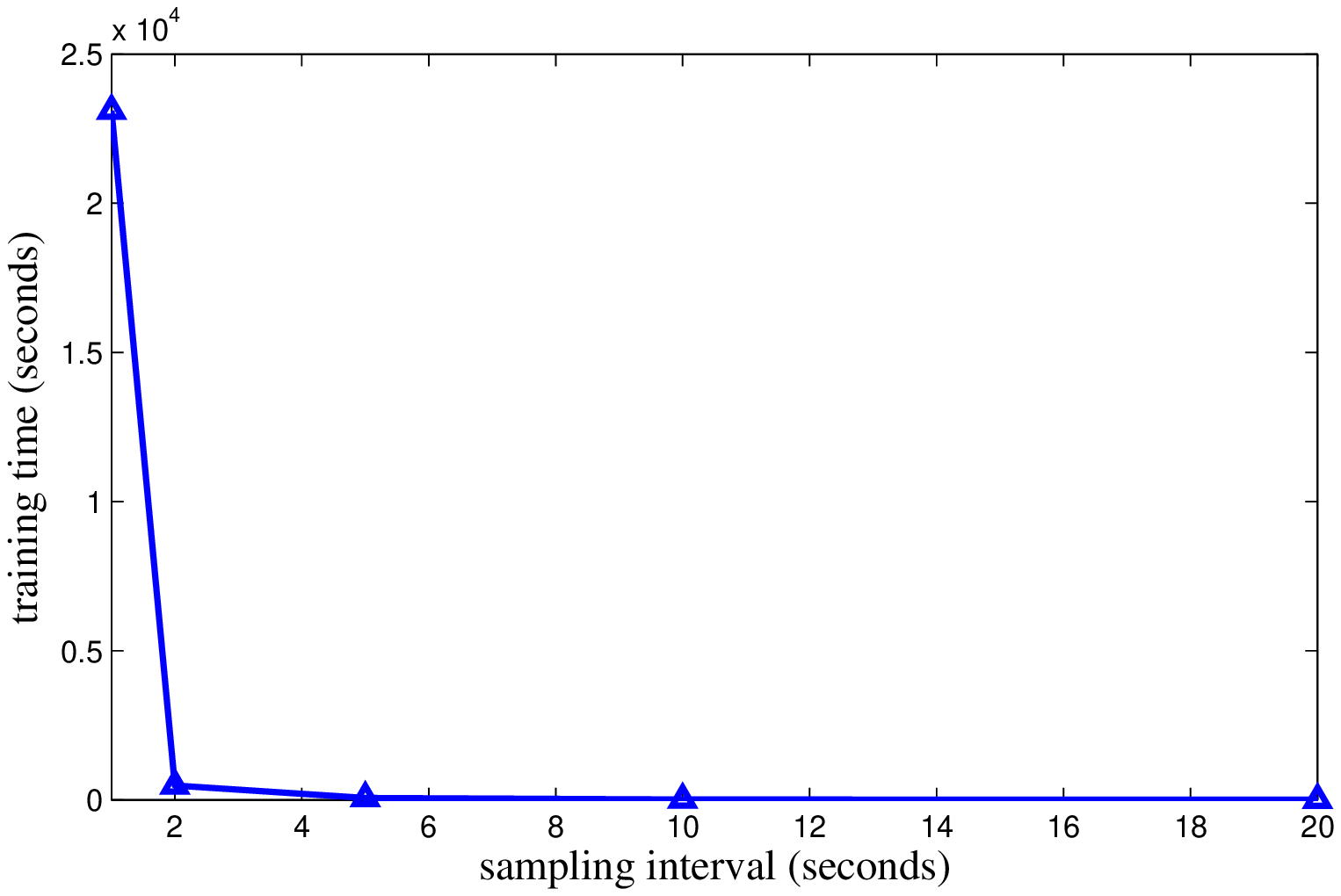}}
\subfigure[]{
\label{time_GCU} 
\includegraphics[width=3.05in,height=2.15in]{time_size_GCU_in.eps}}

\caption{(a),(b) represent respectively the time for training a user's profile by using the SVM algorithm for three-sensors-based system in the PU data set and the GCU data set. (c),(d) zoom in on (a),(b) to the sampling interval from 1 to 20 seconds for clarity. We can see that the smaller sampling interval (high sampling rate) needs more time to train a user's profile. Therefore, we need to find a trade-off sampling rate to balance performance and complexity.}
\label{time} 
\end{figure*}

\begin{table*}[!t]
\caption{Time for training a user's profile by using the SVM algorithm for three sensors, for (a) the PU data set and (b) the GCU data set, respectively. We can see that the smaller sample interval (higher sampling rate) needs more time to train a user's profile. Therefore, we need to find a trade-off sampling rate to balance performance and complexity.}
\begin{tabular}{|c|c|c|c|c|c|c|c|}
\hline
\tabincell{c}{sampling interval} &1&2&5& 10 & 20 & 40 & 60 \\ \hline
\tabincell{c}{training time (PU data set)} & 33502s & 1855s & 170.72s & 39.85s & 6.07s & 1.19s & 0.51s\\ \hline
training time (GCU Data Set) & 23101s & 485s & 62.41s & 9.43s & 1.02s & 0.21s & 0.17s\\ \hline
\end{tabular}
\label{table2}
\end{table*} 
 \begin{figure*}[!t]
\centering
\subfigure[PU data set]{
\label{time_PU} 
\includegraphics[width=3.05in,height=2.15in]{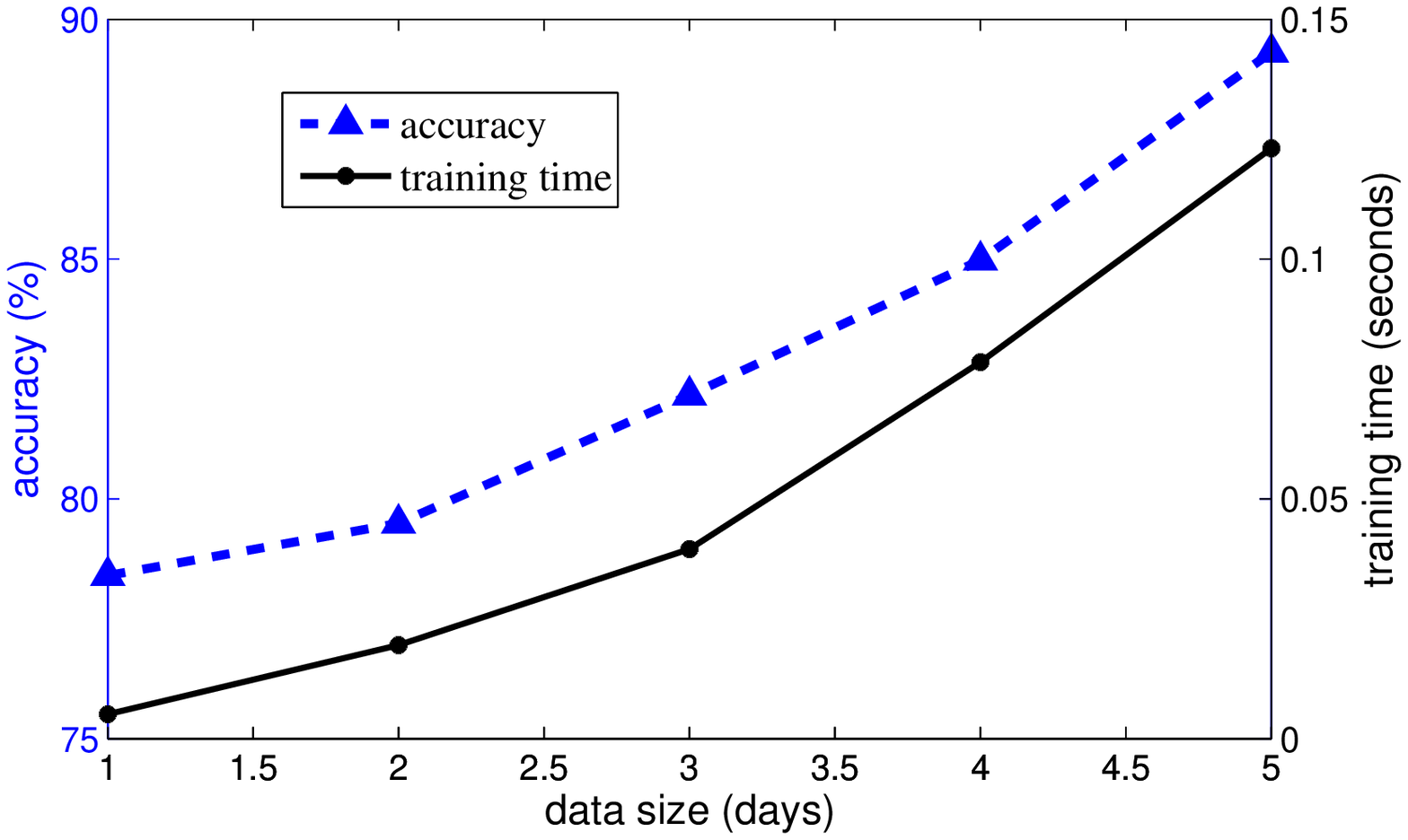}}
\subfigure[GCU data set]{
\label{time_GCU} 
\includegraphics[width=3.05in,height=2.15in]{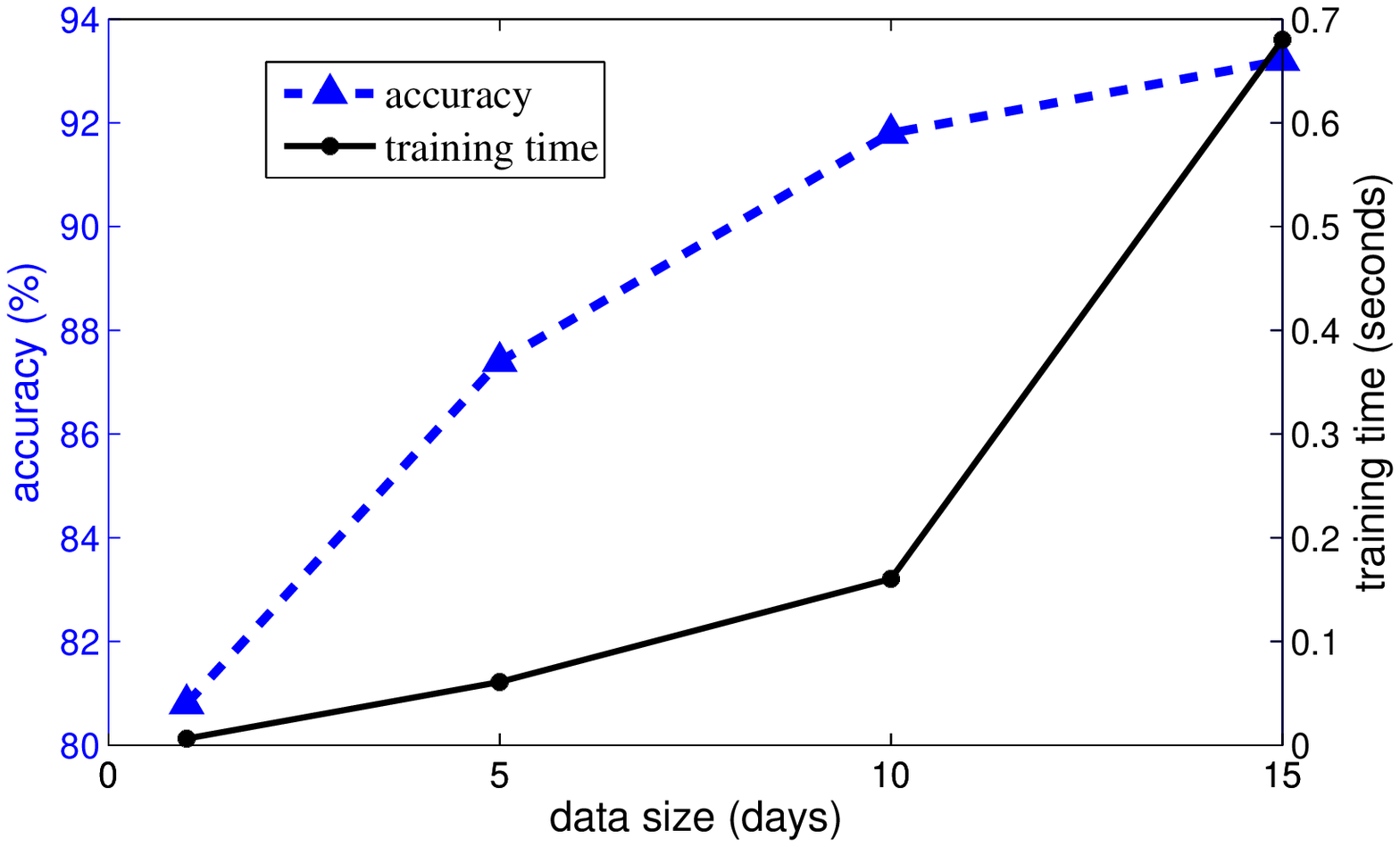}}
\caption{(a),(b) represent the authentication accuracy and the training time with different training data size for three sensors in the PU data set and the GCU data set. Blue dashed lines show that the larger data size has better accuracy because we use more information about the user. Black solid lines show that larger data size usually needs longer training time.}
\label{fig3} 
\end{figure*}

\subsection{Three-sensor authentication}
\begin{spacing}{1}
Now, we compare the three-sensor-based system with one and two sensor-based authentication experiments. From Figure \ref{all} and Table \ref{num}, we observe that the three-sensor results give the best authentication accuracy, as represented by the top line with triangles in both data sets, seen more clearly as the highest value in each column in Table \ref{num}.  Again, we find that the accuracy increases with faster sampling rates because we use more detailed information from each sensor.
\subsection{Training Time vs. Sampling Rate} 
In the rest of the evaluations below, we use the three-sensor-based system, since it has the best authentication accuracy.

From Figure \ref{all} and Table \ref{num}, when the sampling rate is higher than 4 minutes (samples every 240 seconds or less), the accuracy in the PU data set is better than 80\%, while that in the GCU data set is better than 90\%. The average improvement from two sensors to three sensors is 3.3\% in PU data set (4.4\% in GCU data set) when the sampling rate is 20 seconds. Furthermore, when the sampling rate is higher than 20 seconds, the accuracy in the PU data set is better than 90\%, while that in the GCU data set is better than 95\%.

Figure \ref{time} and table \ref{table2} shows that a higher sampling rate (smaller sampling interval) needs more time to train a user's profile. The time exponentially increases with the increase of the sampling rate. It is a trade-off between security and convenience. However, the good news is that when the sampling interval is about 20 seconds, it only needs less than 10 seconds in the PU data set (and roughly 1 second in the GCU data set) to train a user's profile, but the accuracy is higher than 90\% (and 95\% in the GCU data set), as seen from Table \ref{all}. It means that a user only needs to spend less than 10 seconds to train a new model to do the implicit authentication for the whole day in the PU data set and only 1 second for the GCU data set. 

These findings validate the effectiveness of our method and its feasibility for real-world applications. Furthermore, our method can be customized for users. They can change their security level by easily changing the sampling rate of sensors in their smartphones.
\subsection{Accuracy and Time vs. Data size}
Figure \ref{fig3} shows another trade-off between security and convenience. We choose a sampling interval of 10 minutes and a training data size ranging from 1 day to 5 days in the PU data set (and 1 day to 15 days in the GCU data set). The blue dashed line with triangles shows that the accuracy increases with the increase of training data size. The black solid line with circles shows that the training time increases with the increase of training data size.
\end{spacing}
\begin{spacing}{0.5}
\section{CONCLUSION}
\end{spacing}
\noindent
In this paper, we utilize three sensors: the accelerometer, the orientation sensor and the magnetometer. We apply the SVM technique as the classification algorithm in the system, to distinguish the smartphone's owner versus other users, who may potentially be attackers or thieves. In our experiments, we compare the authentication results for different sampling rates and different data sizes, which show a trade-off between accuracy performance and the computational complexity. Furthermore, we experiment with data from a single sensor and from a combination of two sensors, to compare their results with data from all three sensors. We find that the authentication accuracy for the orientation sensor degrades more than that of the other two sensors. Therefore, the data collected from the orientation sensor is not as important as that from the accelerometer and magnetometer, which tend to measure more stable, longer-term characteristics of the user's coarse-grained movements and his general physical location, respectively.

Utilizing sensors to do implicit user authentication is very interesting and promising. Our work also suggests some other interesting research directions. First, we can use more detailed sensors' information to further improve the authentication accuracy. Second, we can try to combine the time information with frequency information to better obtain a user's profile. Many other issues relating to the user's privacy remain. It is also interesting to launch an attack through the sensors' information because our research also shows that indeed, sensors can represent a user's characteristic behavior and physical environment.
\section*{\uppercase{Acknowledgements}}

\noindent 
This work was supported in part by NSF CNS-1218817. We also appreciate the help of Dr. Gunes Kayacik at Glasgow Caledonian University for providing us the GCU data set.
\vfill
\bibliographystyle{apalike}
{\small
\bibliography{v3}}
\end{document}